# A Simple Model For Scalar Relativistic Corrections To Molecular Total Atomization Energies

Jan M.L. Martin[1, a)] and Nitai Sylvetsky[1]

[1] *Department of Organic Chemistry, Weizmann Institute of Science, 76100 Reḥovot, Israel*

a)Corresponding author: gershom@weizmann.ac.il

**Abstract.** Scalar relativistic corrections to atomization energies of 1st-and 2nd-row molecules can be rationalized in terms of a simple additive model, linear in changes in atomic *s* populations. In a sample of 200 first-and second-row molecules, such a model can account for over 98% of the variance (99% for the first-row subset). The remaining error can be halved again by adding a term involving the change in atomic *p* populations: those coefficients need not be fitted but can be fixed from atomic electron affinity calculations. This model allows a fairly accurate *a priori* estimate for the importance of scalar relativistic corrections on a reaction energy, at essentially zero computational cost. While this is not a substitute for explicit calculation of Douglas-Kroll-Hess (DKH) or exact two-component (X2C) relativistic corrections, the model offers an interpretative tool for the chemical analysis of scalar relativistic contributions to reaction energies.

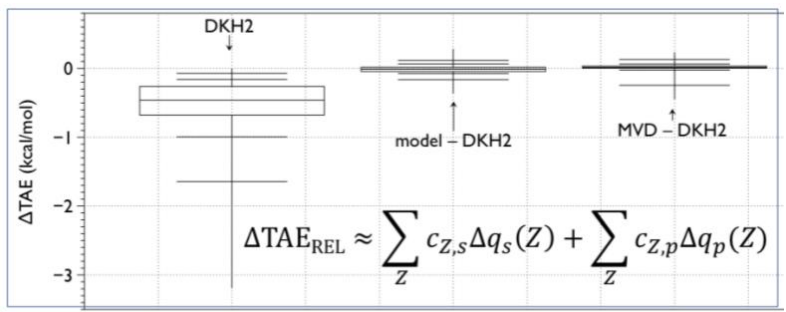

(Mol. Phys. Manuscript TMPH-2018-0265.R1)

## INTRODUCTION

The importance of relativistic effects in the chemistry of heavy elements is fairly universally known among theoretical chemists: for recent reviews, see Pyykkö[1,2] (see also these older reviews[3,4] as well as two textbooks on relativistic quantum chemistry[5,6]).

Less commonly appreciated is that, for accurate thermochemistry, relativistic effects need to be accounted for even in the first two rows of the Periodic Table. For instance, in 1999, it was shown[7] that scalar relativistic effects reduce the atomization energy of $BF_3$ — a key species for fixing the heat of formation of gaseous boron[8,9] — by 0.7 kcal/mol. Perhaps the first convincing evidence that such corrections could affect light elements was the pioneering work by Garcia de la Vega on atomic electron affinities at the Hartree-Fock level[10,11] — extended later[12] through basis set limit full CI extrapolation, and shown there unambiguously to be the 'missing link' in achieving millielectronvolt accuracy. (See also Ref.[13].)

In the computational thermochemistry community — particularly when 1 kJ/mol (0.24 kcal/mol) accuracy is aimed for — the inclusion of scalar relativistic corrections has since become common. We cite here such standardized protocols as the Weizmann-n approaches[14–16] W1, W2, and W4[17,18] developed at the Weizmann Institute, their explicitly correlated Wn-F12 variants,[19–22] the HEAT protocols developed by an international consortium around John F. Stanton,[23–25] and the ccCA approach of Wilson.[26–28] Likewise, in the more general FPD approach,[29–33] scalar relativistic corrections are a standard step.

Our recent W4-17 benchmark[34] — itself an expanded version of the earlier W4-11 dataset[35] — offers an energy decomposition of 200 accurate total atomization energies as Table S-2 in its supporting information. The scalar relativistic components, specifically, were evaluated using the second order Douglas-Kroll approach (DKH2)[36–38] at the CCSD(T)/AV(Q+d)Z level. Some individual cases exceed 2 kcal/mol — such as $SF_6$ (–3.19 kcal/mol), $HClO_4$ (–2.72 kcal/mol), and $PF_5$ (–2.60 kcal/mol). Moreover, benzene, $N_2O_4$, and beta-lactim are just a few of the *first*-row molecules for which corrections reach or exceed –1 kcal/mol. A box-and-whiskers plot of the data distribution is given in Figure 1.

**FIGURE 1.** Box plot of DKH2-CCSD(T)/aug-cc-pV(Q+d)Z scalar relativistic corrections (kcal/mol) to the total atomization energies in the 200-molecule W4-17 dataset, as well as for first-row and second-row subsets. The outer fences encompass the middle 95% of the distribution, the inner fences 80%, the box 50%.

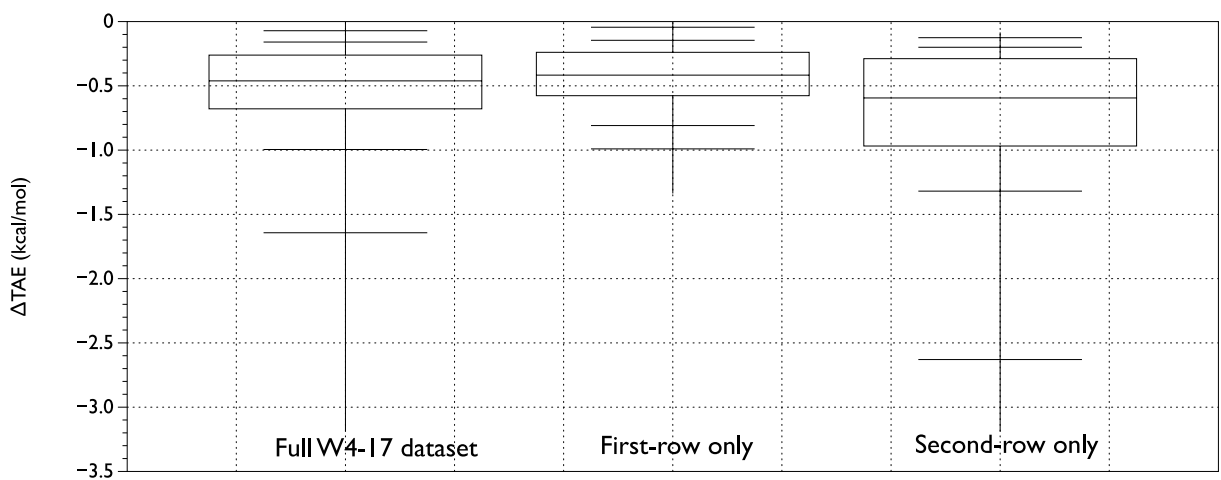

Accounting for these contributions is computationally relatively inexpensive, as DKH2 is sufficient in this accuracy range for the first two rows of the Periodic Table. But as the saying attributed to Eugene P. Wigner goes, "It is nice to know that the computer understands the problem. But I would like to understand it too."

From considering the solutions of the Dirac equation for the hydrogen-like atom, it is clear that relativistic corrections are largest for *s* orbitals, followed by $p_{1/2}$ spinors. It has hence been received wisdom in the relativistic quantum chemical community (see, e.g., Fröman[39,40] for an early example) that changes in *s* orbital population drive many relativistic effects. For instance, the following statement in Dyall et al.[41] comes to mind:

> "The central atom in each of [$BF_3$, $AlF_3$, and $GaF_3$] has a sizeable change in the *s* populations from the atom to the molecule, and hence an appreciable scalar relativistic contribution to the atomization energy."

(Compare also p. 458 of Dyall and Faegri,[5] and Section 16.1 of Reiher and Wolf.[6]) It occurred to us that it would be helpful to verify this conjecture for a significantly-sized sample of molecular data. We will use here the W4-17 dataset, and we will show that not only is there a clear statistical link with the atomic *s* populations, but that a simple additive model based on computed changes in *s* populations can account for over 98% of the variance in the dataset, and over 99% for first-row molecules. By adding correction terms for *p* populations, 99% of the variance in the dataset can be recovered for both first-and second-row molecules.

There has been at least one past attempt to construct a simple model for scalar relativistic corrections (and core-valence corrections) [Ref.[42], eq. (11) and Table V there]: It was based on bond orders and simple multipliers. The present model has more flexibility and predictive power, while requiring a similar number of adjustable parameters.

## COMPUTATIONAL METHODS

Nearly all calculations were performed using the MOLPRO 2015.1 program system[43] running on the Faculty of Chemistry HPC facility. Population analyses were obtained using Gaussian 09,[44] on the same platform.

All reference geometries were taken from the ESI of Ref. [34] Scalar relativistic corrections were recalculated at the CCSD(T) level,[45,46] i.e., coupled cluster[47] with all single and double substitutions plus a quasiperturbative account for connected triple excitations, using the 2nd- order Douglas-Kroll-Hess (DKH2) approach, [36–38] as well as using the "exact two-component" (X2C) approach,[48] using the aug′-cc-pV(Q+d)Z basis set[49–52] for the nonrelativistic energy and de Jong et al.'s relativistic recontraction[53] of this basis set for the relativistic energy.† As additional data, we calculated the 1st-order mass-velocity and Darwin corrections[54,55] as expectation values at the AQCC (averaged quadratic coupled cluster[56]) level, and compared them with both DKH2 and X2C corrections at the same level of theory. Hartree-Fock level corrections were obtained as by-products.

We used two different approaches to assess the occupation of atomic $s$, $p$, and (for 2nd-row atoms in high oxidation states) $d$ orbitals in the molecules. (Both were obtained from PBE0 densities[57,58] using the aug-cc-pV(T+d)Z basis set.[52]) The first is the widely used NPA (natural population analysis[59]) of Weinhold and coworkers, as implemented in the NBO 6 program[60] (although this specific feature is also available in the older NBO 3, which is built into many electronic structure programs). The second is "minimal basis set projected Mulliken" (MBS-Mulliken) as proposed by Montgomery et al.[61] in an attempt to eliminate the pathological basis set dependence of the original Mulliken population analysis. In MBS-Mulliken, the converged orbitals are first projected to an STO-3G(*) basis set (the star refers to the addition of 3d functions on 2nd-row atoms), and a Mulliken population analysis is carried out on those orbitals. The resulting populations are only weakly dependent on the original basis set, and thus satisfy Cioslowski and Surjan's weakened observability criterion.[62]

Multivariate linear regression was carried out using the RegressIt plugin for Microsoft Excel 2016 (http://www.regressit.com, Macintosh version) as well as using the built-in Solver functionality of Excel.

## RESULTS AND DISCUSSION

Table 1 presents the parameters and statistics of various models fit to the data. We consider statistics both for the 129 first-row only molecules, and for the complete set of 200.

First, we consider the DKH2 correction at the Hartree-Fock level and the simple model

$$\Delta \text{REL} = \sum_Z c_{Z,s} \Delta q_s(Z)$$

in which the sum of partial charge changes for a given element $Z$ and angular momentum $l$ is given by:

$$\Delta q_l(Z) = \sum_{Z_i=Z} q_{l,i}(\text{molecule}) - q_{l,i}(\text{atom})$$

Despite its simplistic nature, it recovers an astonishing 99.7% of the sum of squares (SSQ) of the first-row data, whether using NPA or MBS-Mulliken charges. These numbers deteriorate somewhat upon the introduction of electron correlation — this is not unexpected for molecules with significant static correlation, where some of the prominent excited determinants may have charge distributions very different from the Hartree-Fock ground state. Yet still over 99% of the SSQ is being described, which is no mean feat considering the primitive nature of the model and its having just six adjustable parameters (one per element H, B–F).

While the fitted parameters for NPA and MBS-Mulliken charges differ, the quality of the fits is comparable.

---

† In the process, we found that the W4-17 dataset contains a typo: the DKH2 scalar relativistic correction for PF5 in the ESI of Ref. [34] should read –2.57 kcal/mol rather than –3.52 kcal/mol (which was obtained using a smaller basis set). We thank Prof. Amir Karton (U. of Western Australia) for clarifying this.

**TABLE 1.** Sum of squares of the data and the residual error calculated by different fitted models

| Method | Fitted data | $N_{param}$ | $N_{data}$ | $SSQ_{data}$ | $SSQ_{residue}$ | %residue | %$SSQ_{data}$ recovered |
|---|---|---|---|---|---|---|---|
| *PBEO/AV(T+d)Z MBS-Mulliken* | | | | | | | |
| DKH2 CCSD(T) | row1 | 6 | 129 | **33.82** | **0.29** | **0.87** | **99.1** |
| DKH2 CCSD(T) | row1(frozen), row2 | 5 | 200 | **97.95** | **1.73** | **1.77** | **98.2** |
| DKH2 SCF | row1 + row2 | 11 | 200 | **97.95** | **1.64** | **1.67** | **98.3** |
| DKH2 SCF | row1(frozen), row2 + $p$ | 10 | 200 | **98.95** | **1.18** | **1.19** | **98.8** |
| DKH2 SCF | row1(frozen), row2 + $d$ | 10 | 200 | **97.95** | **1.61** | **1.64** | **98.4** |
| DKH2 SCF | row1(frozen), row2 + $p$ + $d$ | 15 | 200 | **97.95** | **1.08** | **1.11** | **98.9** |
| DKH2 SCF | row1 | 6 | 129 | **54.01** | **0.18** | **0.33** | **99.7** |
| DKH2 SCF | row1(frozen), row2 | 5 | 200 | 163.78 | 2.87 | 1.75 | 98.2 |
| *PBEO/AV(T+d)Z NBO* | | | | | | | |
| DKH2 CCSD(T) | row1 | 6 | 129 | **33.86** | **0.25** | **0.75** | **99.3** |
| DKH2 CCSD(T) | row1(frozen), row2 | 5 | 200 | **97.95** | **4.68** | **4.78** | **95.2** |
| DKH2 CCSD(T) | row1(frozen), row2 [a] | 5 | 198 | **80.34** | **2.00** | **2.49** | **97.5** |
| DKH2 SCF | row1 + row2 | 11 | 198 | **80.34** | **1.86** | **2.32** | **97.7** |
| DKH2 SCF | row1(frozen), row2 + $d$ | 10 | 200 | **97.95** | **2.93** | **3.00** | **97.0** |
| DKH2 SCF | row1(frozen), row2 + $p$ + $d$ | 15 | 200 | **97.95** | **1.88** | **1.92** | **98.1** |
| DKH2 SCF | row1 | 6 | 129 | **54.01** | **0.18** | **0.33** | **99.7** |
| DKH2 SCF | row1(frozen), row2 | 5 | 200 | **163.78** | **4.57** | **2.79** | **97.2** |
| DKH2 SCF | row1(frozen), row2 [a] | 5 | 198 | **137.13** | **2.16** | **1.57** | **98.4** |
| AQCC MVD | row1 | 6 | 129 | **30.97** | **0.42** | **1.34** | **98.7** |
| AQCC MVD | row1(frozen), row2 | 5 | 200 | **102.36** | **7.99** | **7.81** | **92.2** |
| AQCC MVD | row1(frozen), row2 [a] | 5 | 198 | **80.37** | **3.95** | **4.91** | **95.1** |

[a] omitting $SF_6$ and $HClO_4$ as outliers

It is perhaps not surprising that the fits would deteriorate somewhat when we broaden our scope to 2nd-row compounds, i.e., to the full 200-molecule set. Somewhat more intriguingly, however, a rift opens between the NPA and MBS-Mulliken based models. Using MBS-Mulliken charges, 98.2% of SSQ is still recovered, compared to just 95.2% for NPA charges. The latter value can be improved to 97.5% by eliminating two severe outliers, namely, the pseudohypervalent compounds $HClO_4$ and $SF_6$. No similar outliers are seen in the MBS-Mulliken case.

The fitted parameters for H and B–F, both for MBS-Mulliken and for the NPA fit without the two outliers, are fairly close to those obtained from the first-row fit. Indeed, simply adopting the 1st-row values and refitting just the Al–Cl parameters causes only a marginal degradation of SSQ.

We attempted introducing additional parameters for $\Delta q_p(Z)$. For the first row, these cause no noticeable improvement in the fit, while for the second row we do see one, e.g., with the MBS-Mulliken charges, adding p coefficients for Al–Cl reduces the SSQ residual from 1.77% to 1.19%, with a further reduction to 1.10% from adding a $\Delta q_d(Z)$ term with $d$ coefficients for P, S, Cl. While these improvements do pass the Fisher-Snedecor test for statistical significance, particularly the added $d$ coefficients may amount to 'gilding the lily'. For the NBO-based model, the improvements for adding $p$ and $d$ coefficients are more significant, at least when the outliers are included: from 4.78% to 3.00% to 1.92%.

Considering that relativistic corrections for $d$ orbitals would be smaller still than those for $p$ orbitals, a statistically significant improvement from $d$ coefficients for 2nd-row atoms would seem counterintuitive. However, one should keep in mind that essentially the only 2nd-row molecules for which those might matter are pseudohypervalent ones[63] — exactly the ones with which the NBO-based model struggles.

Still, one might argue that any attempt to squeeze more than 98% or so of variance out of such a simplistic model amounts to an exercise in 'kitchen sink regression', and that for any higher accuracy, people should just carry out actual relativistic calculations rather than rely on a semiempirical estimate like the present one. What the latter does offer is a semiquantitative *a priori* estimate of the importance of scalar relativistic corrections.

There is, alas, one notable, clear-cut relativistic effect that a model based on only $\Delta q_s(Z)$ is intrinsically unable to capture: the scalar relativistic correction to atomic electron affinities[10–12] (which for B-F and Al-Cl basically corresponds to the effect of an extra p electron). Since the $p$ parameters for the 1st row are ill-determined statistically, one could instead assign fixed values from the calculated contributions to electron affinities, ionization potentials, or their average — the latter corresponds[64] to the Mulliken electronegativity except for a constant.

We could now, of course, assign fixed values $c_{Z,p}=\Delta EA_{rel}$ to the coefficients in the $p$ term:

$$\Delta \text{REL} = \sum_Z c_{Z,s} \Delta q_s(Z) + \sum_Z c_{Z,p} \Delta q_p(Z)$$

If we do so, and: (a) fit first the parameters for H and B-F to MBS-Mulliken charges the 129 first-row species, then: (b) freeze those and fit the parameters for the 2$^{nd}$-row atoms to the remaining 70 points (SF$_6$ was once again found to be an outlier), we find

$$\begin{aligned}\Delta \text{REL} = {}& 0.137(15).\Delta q_s(H) + 0.051(13).\Delta q_s(B) + 0.114(4).\Delta q_s(C) + 0.240(15).\Delta q_s(N) + 0.653(28).\Delta q_s(O) \\ & + 1.489(60).\Delta q_s(F) + 0.658(30).\Delta q_s(Al) + 0.941(28).\Delta q_s(Si) + 1.815(52).\Delta q_s(P) \\ & + 2.795(61).\Delta q_s(S) + 4.984(84).\Delta q_s(Cl) + \Delta \text{REL}_p\end{aligned}$$

in which the uncertainties in parameters represent 95% confidence intervals, and the fixed $p$ corrections are

$$\begin{aligned}\Delta \text{REL}_p = {}& -0.028.\Delta q_p(B) - 0.063.\Delta q_p(C) - 0.082.\Delta q_p(N) - 0.140.\Delta q_p(O) - 0.221.\Delta q_p(F) \\ & -0.117.\Delta q_p(Al) - 0.173.\Delta q_p(Si) - 0.199.\Delta q_p(P) - 0.261.\Delta q_p(S) - 0.324.\Delta q_p(Cl)\end{aligned}$$

As can be seen in Table 2, this model captures 99.3% of variance for the first row, and 99.0% for both rows (with SF$_6$ omitted). It should be noted that the ratio between $s$ and $p$ coefficients for a given element grows with Z from about a factor of two to fifteen: hence, clamping the $p$ coefficients rather than fitting them does not greatly impact accuracy.

As also shown in Table 2, the SSQ recovery for both rows can be increased to 99.2% by adding $d$ parameters; if the outlier point SF$_6$ is brought back in, we are back at 98.7%. These parameters are not statistically well-determined, however, and we have chosen to omit them in deference to Occam's law of parsimony.

With NBO-based charges, things are not as simple. We can recover 99.5% for the first row, but for both rows together, this drops to just 92.2% — deleting $HClO_4$ and $SF_6$ as outliers brings us back to 95.2%, which can be lifted up to 98.0% even with $HClO_4$ included if a $\Delta q_d(Z)$ term is added.

**TABLE 2.** Sum of squares of the data and the residual error calculated by fitted models to which EA-derived $p$ parameters were added

| Method | Fitted data | $N_{param}$ | $N_{data}$ | $SSQ_{data}$ | $SSQ_{residue}$ | %residue | %$SSQ_{data}$ recovered |
|---|---|---|---|---|---|---|---|
| | PBE0/AV(T+d)Z MBS-Mulliken + p(EA) | | | | | | |
| DKH2 CCSD(T) | row1 | 6 | 129 | **33.82** | **0.24** | **0.70** | **99.3** |
| | row1(frozen), row2 | 5 | 200 | **97.95** | **1.56** | **1.59** | **98.4** |
| | row1(frozen), row2 [a] | 5 | 199 | **87.77** | **0.85** | **0.97** | **99.0** |
| | row1(frozen), row2 + d [a] | 10 | 199 | **87.77** | **0.67** | **0.77** | **99.2** |
| | row1(frozen), row2 + d | 10 | 200 | **97.95** | **1.24** | **1.26** | **98.7** |
| | row1 + row2 | 11 | 200 | **98.95** | **1.12** | **1.13** | **98.9** |
| | PBE0/AV(T+d)Z NBO + p(EA) | | | | | | |
| DKH2 CCSD(T) | row1 | 6 | 129 | **33.86** | **0.16** | **0.48** | **99.5** |
| | row1(frozen), row2 | 5 | 200 | **97.95** | **7.67** | **7.83** | **92.2** |
| | row1(frozen), row2 [b] | 5 | 198 | **80.34** | **3.89** | **4.85** | **95.2** |
| | row1(frozen), row2 + d [a] | 10 | 199 | **87.77** | **1.73** | **1.97** | **98.0** |
| | row1 + row2 | 11 | 200 | **87.77** | **1.73** | **1.98** | **98.0** |

[a] omitting $SF_6$ as an outlier
[b] omitting $SF_6$ and $HClO_4$ as outliers

**FIGURE 2.** Box plot of differences (kcal/mol) between our model and the reference DKH2-CCSD(T)/aug-cc-pV(Q+d)Z scalar relativistic corrections for the whole W4-17 dataset, as well as for first-row and second-row subsets. Differences between MVD-AQCC/aug-cc-pV(Q+d)Z data and the reference are also plotted. The outer fences encompass the middle 95% of the distribution, the inner fences 80%, the box 50%.

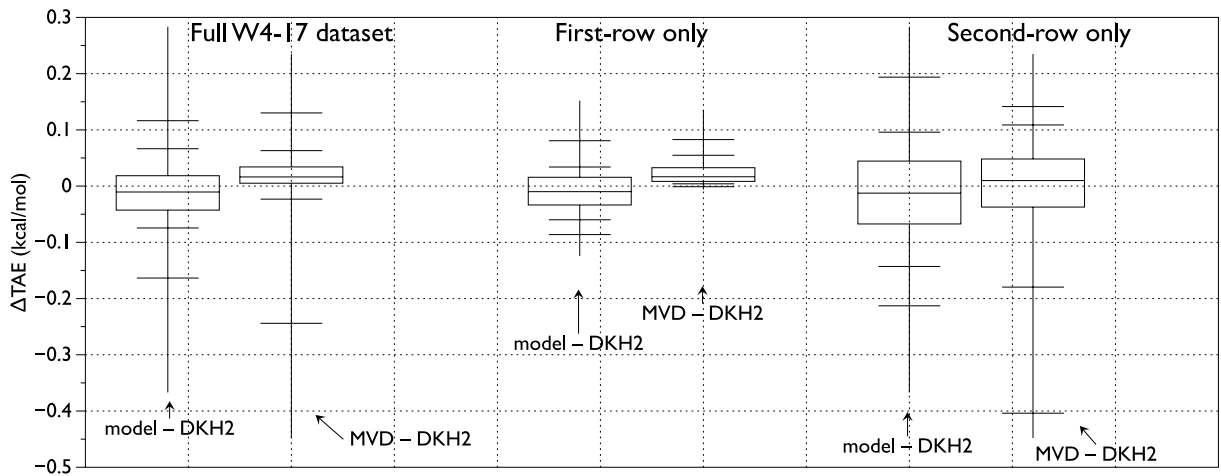

The accuracy of our model in terms of actual errors for atomization energies is represented as a box-and-whiskers plot in Figure 2, which can be compared with Figure 1 for the magnitude of the DKH2 corrections. Statistics, including interquartile (IQR) and interdecile (IDR) ranges of the data distribution, are given in Table 3, together with selected values for individual molecules.

Errors in the model are between ±0.07 kcal/mol for 80% of the W4-17 sample, and [–0.16, +0.12] kcal/mol for 95% of the sample; for the first-row subset, this latter interval shrinks to [–0.09,0.08] kcal/mol, compared to a 95% spread in the scalar relativistic corrections themselves of [-0.99,-0.04] kcal/mol; for the second-row subset, the 95% interval of the model errors broadens to [-0.21,0.19] kcal/mol, compared to a 95% spread of [-2.63,-0.13] kcal/mol for the actual values.

**TABLE 3.** DKH2-CCSD(T)/aug-cc-pV(Q+d)Z scalar relativistic corrections to total atomization energies (kcal/mol) for selected first-and second-row molecules, and errors at the MVD-AQCC/aug-cc-pV(Q+d)Z level as well as the final ΔREL partial-changes model. Averages and standard deviations for the whole sample as well as for first-and second-row subsets are also given.

|  | DKH2 reference | Model–DKH2 | MVD–DKH2 |  | DKH2 reference | Model–DKH2 | MVD–DKH2 |
|---|---|---|---|---|---|---|---|
| all of W4-17 (200 molecules) | | | | | | | |
| average | -0.543 | -0.012 | 0.010 | | | | |
| RMS | 0.700 | 0.073 | 0.082 | | | | |
| median | -0.461 | -0.010 | 0.016 | | | | |
| IQR (50% interval) | 0.414 | 0.061 | 0.029 | | | | |
| IDR (80% interval) | 0.836 | 0.141 | 0.086 | | | | |
| 95% interval | 1.572 | 0.280 | 0.374 | | | | |
| First-row only (129 molecules) | | | | Second-row only (71 molecules) | | | |
| average | -0.438 | -0.009 | 0.024 | average | -0.733 | -0.019 | -0.016 |
| RMS | 0.512 | 0.073 | 0.034 | RMS | 0.950 | 0.108 | 0.129 |
| median | -0.417 | -0.010 | 0.017 | median | -0.594 | -0.013 | 0.010 |
| IQR (50% interval) | 0.338 | 0.061 | 0.024 | IQR (50% interval) | 0.679 | 0.112 | 0.085 |
| IDR (80% interval) | 0.663 | 0.141 | 0.051 | IDR (80% interval) | 1.119 | 0.239 | 0.288 |
| 95% interval | 0.947 | 0.280 | 0.084 | 95% interval | 2.504 | 0.407 | 0.545 |
| Selected examples | | | | Selected examples | | | |
| Benzene | -0.995 | -0.037 | 0.011 | $HClO_4$ | -2.724 | -0.051 | -0.440 |
| $C_2F_6$ | -1.337 | 0.040 | 0.063 | $C_2Cl_6$ | -1.128 | -0.139 | -0.180 |
| $N_2O_4$ | -0.996 | 0.068 | 0.136 | $AlCl_3$ | -1.284 | 0.283 | 0.172 |
| Dioxetane | -0.844 | 0.152 | 0.034 | $ClF_5$ | -0.711 | 0.227 | -0.448 |
| n-Pentane | -0.958 | 0.070 | 0.014 | $SO_3$ | -1.846 | -0.367 | -0.024 |
| $N_2O$ | -0.455 | -0.011 | 0.057 | $P_4$ | -0.725 | -0.129 | -0.091 |
| $CO_2$ | -0.480 | -0.039 | 0.032 | $SiF_4$ | -1.901 | 0.055 | 0.060 |
| Formamide | -0.647 | 0.011 | 0.022 | Thiophene | -1.064 | -0.064 | 0.024 |
| Tetrahedrane | -0.776 | 0.032 | 0.011 | $PF_5$ | -2.598 | 0.098 | -0.075 |
| Acetic acid | -0.790 | -0.002 | 0.028 | $Si_2H_6$ | -1.319 | -0.036 | 0.235 |

In nonparametric statistics, the interquartile range (IQR) is defined as the distance between the 25[th] and 75[th] percentile of the data distribution, and the interdecile range (IDR) as the distance between the 10[th] and 90[th] percentile. The range between the 2.5[th] and 97.5[th] percentile has also been added.

For perspective, let us compare with the difference between calculated DKH2 corrections and more approximate 1st-order Darwin and mass-velocity corrections. While 1st-order MVD generally performs quite well for 1st-row compounds (median difference of 0.03 kcal/mol), discrepancies of up to 0.2 kcal/mol are seen for some individual cases. In the second row, 1st-order MVD performance is more erratic, discrepancies with DKH2 reaching up to 0.35 and -0.25 kcal/mol. Especially at the margins of the distribution (Figure 2), the errors in our model are in fact comparable with the differences between MVD and DKH2. This does *not* mean, however, that we recommend replacing DKH2 calculations by our model: instead, we suggest the latter as a tool for chemically rationalizing the scalar relativistic corrections and for predicting whether a reaction energy is likely to have a nontrivial such contribution.

The example molecules in Table 3 illustrate that, while the model overall works better than could be reasonably expected given its arguably simplistic nature, remaining errors are quite unsystematic, and the model is no substitute for calculation.

Finally, let us consider the difference between calculated DKH2 corrections and more rigorous X2C (exact 2-component) values. For the first-row subset of W4-17, the DKH2 and X2C corrections are for thermochemical purposes indistinguishable; for second-row, the largest differences between them are on the order of 0.01 kcal/mol (for $C_2Cl_6$). We conclude that DKH2 is converged in terms of the scalar relativistic treatment for elements lighter than argon. Generally speaking, we recommend eschewing MVD in favor of DKH2 (which is available in most major electronic structure codes) or, where available, X2C, which is functionally equivalent for the systems studied here but may be more robust for heavier elements.

## CONCLUSIONS

We have shown here that scalar relativistic corrections to atomization energies of 1st-and 2nd-row molecules can be rationalized neatly by a simple additive model in terms of changes in atomic *s* populations. Such a model can account for over 99% of the variance for 1st-row molecules, and about 98% for 1st and 2nd-row molecules together: the latter statistic can be improved further by adding a term involving the change in atomic *p* populations, but fixing its parameters to values obtained from atomic electron affinity calculations. While the said model is not a substitute for proper DKH2 or X2C calculations, it allows a fairly accurate *a priori* estimate for the importance of scalar relativistic corrections on a reaction energy, as well as a means for interpreting actual calculated scalar relativistic corrections.

## ACKNOWLEDGMENTS

This research was supported by the Israel Science Foundation (grant 1358/15), the Minerva Foundation, and the Helen and Martin Kimmel Center for Molecular Design (Weizmann Institute of Science). The authors would like to thank Dr. Kenneth G. Dyall (Schrödinger, Inc.) for enlightening discussions at the 58th Sanibel Symposium, and Prof. Pekka Pyykkö (U. of Helsinki) for helpful comments and for bringing the early work of Anders Fröman[39,40] to our attention. This paper is dedicated to the memory of Prof. Israel Rubinstein OBM (1947-2017), outstanding scientist, friend, and Renaissance man.

## SUPPORTING INFORMATION

The computed and model-generated scalar relativistic corrections for the W4-17 dataset, as well as the population analyses, are freely available online in the FigShare data repository at http://doi.org/10.6084/m9.figshare.6154418, reference number 6154418.

## REFERENCES

(1)   Pyykkö, P. The Physics behind Chemistry and the Periodic Table. *Chem. Rev.* **2012**, *112* (1), 371–384 DOI: 10.1021/cr200042e.

(2) Pyykkö, P. Relativistic Effects in Chemistry: More Common Than You Thought. *Annu. Rev. Phys. Chem.* **2012**, *63* (1), 45–64 DOI: 10.1146/annurev-physchem-032511-143755.
(3) Pyykkö, P.; Desclaux, J. P. Relativity and the Periodic System of Elements. *Acc. Chem. Res.* **1979**, *12* (8), 276–281 DOI: 10.1021/ar50140a002.
(4) Pyykkö, P. Relativistic Effects in Structural Chemistry. *Chem. Rev.* **1988**, *88* (3), 563–594 DOI: 10.1021/cr00085a006.
(5) Dyall, K. G.; Faegri, K. *Introduction to Relativistic Quantum Chemistry*; Oxford University Press: New York, 2007.
(6) Reiher, M.; Wolf, A. *Relativistic Quantum Chemistry*; Wiley-VCH Verlag GmbH & Co. KGaA: Weinheim, Germany, 2009.
(7) Bauschlicher, C. W.; Martin, J. M. L.; Taylor, P. R. Boron Heat of Formation Revisited: Relativistic Effects on the BF 3 Atomization Energy. *J. Phys. Chem. A* **1999**, *103* (38), 7715–7718 DOI: 10.1021/jp991713a.
(8) Martin, J. M. L.; Taylor, P. R. Revised Heat of Formation for Gaseous Boron: Basis Set Limit Ab Initio Binding Energies of BF 3 and BF. *J. Phys. Chem. A* **1998**, *102* (18), 2995–2998 DOI: 10.1021/jp9807930.
(9) Karton, A.; Martin, J. M. L. Heats of Formation of Beryllium, Boron, Aluminum, and Silicon Re-Examined by Means of W4 Theory. *J. Phys. Chem. A* **2007**, *111* (26), 5936–5944 DOI: 10.1021/jp071690x.
(10) García de la Vega, J. M. Relativistic Corrections to the Atomic Electron Affinities. *Phys. Rev. A* **1995**, *51* (3), 2616–2618 DOI: 10.1103/PhysRevA.51.2616.
(11) Koga, T.; Aoki, H.; de la Vega, J. M. G.; Tatewaki, H. Atomic Ionization Potentials and Electron Affinities with Relativistic and Mass Corrections. *Theor. Chem. Acc.* **1997**, *96* (4), 248–255 DOI: 10.1007/s002140050227.
(12) de Oliveira, G.; Martin, J. M. L.; de Proft, F.; Geerlings, P. Electron Affinities of the First- and Second-Row Atoms: Benchmark Ab Initio and Density-Functional Calculations. *Phys. Rev. A* **1999**, *60* (2), 1034–1045 DOI: 10.1103/PhysRevA.60.1034.
(13) Sylvetsky, N.; Kesharwani, M. K.; Martin, J. M. L. The Aug-Cc-PVnZ-F12 Basis Set Family: Correlation Consistent Basis Sets for Explicitly Correlated Benchmark Calculations on Anions and Noncovalent Complexes. *J. Chem. Phys.* **2017**, *147* (13), 134106 DOI: 10.1063/1.4998332.
(14) Martin, J. M. L.; de Oliveira, G. Towards Standard Methods for Benchmark Quality Ab Initio Thermochemistry—W1 and W2 Theory. *J. Chem. Phys.* **1999**, *111* (5), 1843–1856 DOI: 10.1063/1.479454.
(15) Parthiban, S.; Martin, J. M. L. Assessment of W1 and W2 Theories for the Computation of Electron Affinities, Ionization Potentials, Heats of Formation, and Proton Affinities. *J. Chem. Phys.* **2001**, *114* (14), 6014–6029 DOI: 10.1063/1.1356014.
(16) Parthiban, S.; Martin, J. M. L. Fully Ab Initio Atomization Energy of Benzene via Weizmann-2 Theory. *J. Chem. Phys.* **2001**, *115* (5), 2051–2054 DOI: 10.1063/1.1385363.
(17) Karton, A.; Rabinovich, E.; Martin, J. M. L.; Ruscic, B. W4 Theory for Computational Thermochemistry: In Pursuit of Confident Sub-KJ/Mol Predictions. *J. Chem. Phys.* **2006**, *125* (14), 144108 DOI: 10.1063/1.2348881.
(18) Karton, A.; Taylor, P. R.; Martin, J. M. L. Basis Set Convergence of Post-CCSD Contributions to Molecular Atomization Energies. *J. Chem. Phys.* **2007**, *127* (6), 064104 DOI: 10.1063/1.2755751.
(19) Karton, A.; Martin, J. M. L. Explicitly Correlated Wn Theory: W1-F12 and W2-F12. *J. Chem. Phys.* **2012**, *136* (12), 124114 DOI: 10.1063/1.3697678.
(20) Chan, B.; Radom, L. W3X: A Cost-Effective Post-CCSD(T) Composite Procedure. *J. Chem. Theory Comput.* **2013**, *9* (11), 4769–4778 DOI: 10.1021/ct4005323.
(21) Chan, B.; Radom, L. W2X and W3X-L: Cost-Effective Approximations to W2 and W4 with KJ Mol −1 Accuracy. *J. Chem. Theory Comput.* **2015**, *11* (5), 2109–2119 DOI: 10.1021/acs.jctc.5b00135.
(22) Sylvetsky, N.; Peterson, K. A.; Karton, A.; Martin, J. M. L. Toward a W4-F12 Approach: Can Explicitly Correlated and Orbital-Based Ab Initio CCSD(T) Limits Be Reconciled? *J. Chem. Phys.* **2016**, *144* (21), 214101 DOI: http://dx.doi.org/10.1063/1.4952410.
(23) Tajti, A.; Szalay, P. G.; Császár, A. G.; Kállay, M.; Gauss, J.; Valeev, E. F.; Flowers, B. A.; Vázquez, J.; Stanton, J. F. HEAT: High Accuracy Extrapolated Ab Initio Thermochemistry. *J. Chem. Phys.* **2004**, *121* (23), 11599–11613 DOI: 10.1063/1.1811608.
(24) Harding, M. E.; Vázquez, J.; Ruscic, B.; Wilson, A. K.; Gauss, J.; Stanton, J. F. High-Accuracy Extrapolated Ab Initio Thermochemistry. III. Additional Improvements and Overview. *J. Chem. Phys.* **2008**, *128* (11), 114111 DOI: 10.1063/1.2835612.
(25) Harding, M. E.; Vázquez, J.; Gauss, J.; Stanton, J. F.; Kállay, M. Towards Highly Accurate Ab Initio

Thermochemistry of Larger Systems: Benzene. *J. Chem. Phys.* **2011**, *135* (4), 044513 DOI: 10.1063/1.3609250.

(26) DeYonker, N. J.; Wilson, B. R.; Pierpont, A. W.; Cundari, T. R.; Wilson, A. K. Towards the Intrinsic Error of the Correlation Consistent Composite Approach (CcCA). *Mol. Phys.* **2009**, *107* (8–12), 1107–1121 DOI: 10.1080/00268970902744359.

(27) DeYonker, N. J.; Cundari, T. R.; Wilson, A. K. The Correlation Consistent Composite Approach (CcCA): Efficient and Pan-Periodic Kinetics and Thermodynamics. In *Advances in the Theory of Atomic and Molecular Systems (Progress in Theoretical Chemistry and Physics, Vol. 19)*; Piecuch, P., Maruani, J., Delgado-Barrio, G., Wilson, S., Eds.; Progress in Theoretical Chemistry and Physics; Springer Netherlands: Dordrecht, 2009; Vol. 19, pp 197–224.

(28) DeYonker, N. J.; Cundari, T. R.; Wilson, A. K. The Correlation Consistent Composite Approach (CcCA): An Alternative to the Gaussian-n Methods. *J. Chem. Phys.* **2006**, *124* (11), 114104 DOI: 10.1063/1.2173988.

(29) Feller, D.; Peterson, K. A.; Dixon, D. A. A Survey of Factors Contributing to Accurate Theoretical Predictions of Atomization Energies and Molecular Structures. *J. Chem. Phys.* **2008**, *129* (20), 204105 DOI: 10.1063/1.3008061.

(30) Li, S.; Hennigan, J. M.; Dixon, D. A.; Peterson, K. A. Accurate Thermochemistry for Transition Metal Oxide Clusters. *J. Phys. Chem. A* **2009**, *113* (27), 7861–7877 DOI: 10.1021/jp810182a.

(31) Bross, D. H.; Hill, J. G.; Werner, H.-J.; Peterson, K. A. Explicitly Correlated Composite Thermochemistry of Transition Metal Species. *J. Chem. Phys.* **2013**, *139* (9), 094302 DOI: 10.1063/1.4818725.

(32) Dixon, D.; Feller, D.; Peterson, K. A Practical Guide to Reliable First Principles Computational Thermochemistry Predictions Across the Periodic Table. *Annu. Rep. Comput. Chem.* **2012**, *8*, 1–28 DOI: 10.1016/B978-0-444-59440-2.00001-6.

(33) Feller, D.; Peterson, K. A.; Ruscic, B. Improved Accuracy Benchmarks of Small Molecules Using Correlation Consistent Basis Sets. *Theor. Chem. Acc.* **2013**, *133* (1), 1407 DOI: 10.1007/s00214-013-1407-z.

(34) Karton, A.; Sylvetsky, N.; Martin, J. M. L. W4-17: A Diverse and High-Confidence Dataset of Atomization Energies for Benchmarking High-Level Electronic Structure Methods. *J. Comput. Chem.* **2017**, *38* (24), 2063–2075 DOI: 10.1002/jcc.24854.

(35) Karton, A.; Daon, S.; Martin, J. M. L. W4-11: A High-Confidence Benchmark Dataset for Computational Thermochemistry Derived from First-Principles W4 Data. *Chem. Phys. Lett.* **2011**, *510*, 165–178 DOI: 10.1016/j.cplett.2011.05.007.

(36) Douglas, M.; Kroll, N. M. Quantum Electrodynamical Corrections to the Fine Structure of Helium. *Ann. Phys. (N. Y).* **1974**, *82* (1), 89–155 DOI: 10.1016/0003-4916(74)90333-9.

(37) Hess, B. A. Relativistic Electronic-Structure Calculations Employing a Two-Component No-Pair Formalism with External-Field Projection Operators. *Phys. Rev. A* **1986**, *33* (6), 3742–3748 DOI: 10.1103/PhysRevA.33.3742.

(38) Reiher, M. Douglas–Kroll–Hess Theory: A Relativistic Electrons-Only Theory for Chemistry. *Theor. Chem. Acc.* **2006**, *116* (1–3), 241–252 DOI: 10.1007/s00214-005-0003-2.

(39) Fröman, A. Correlation Energies of Some He- and Ne-Like Systems. *Phys. Rev.* **1958**, *112* (3), 870–872 DOI: 10.1103/PhysRev.112.870.

(40) Fröman, A. Relativistic Corrections in Many-Electron Systems. *Rev. Mod. Phys.* **1960**, *32* (2), 317–321 DOI: 10.1103/RevModPhys.32.317.

(41) Dyall, K. G.; Bauschlicher, C. W.; Schwenke, D. W.; Pyykkö, P. Is the Lamb Shift Chemically Significant? *Chem. Phys. Lett.* **2001**, *348* (5–6), 497–500 DOI: 10.1016/S0009-2614(01)01162-9.

(42) Martin, J. M. L.; Sundermann, A.; Fast, P. L.; Truhlar, D. G. Thermochemical Analysis of Core Correlation and Scalar Relativistic Effects on Molecular Atomization Energies. *J. Chem. Phys.* **2000**, *113* (4), 1348–1358 DOI: 10.1063/1.481960.

(43) Werner, H.-J.; Knowles, P. J.; Knizia, G.; Manby, F. R.; Schütz, M.; Celani, P.; Korona, T.; Lindh, R.; Mitrushenkov, A.; Rauhut, G.; Shamasundar, K. R.; Adler, T. B.; Amos, R. D.; Bernhardsson, A.; Berning, A.; Cooper, D. L.; Deegan, M. J. O.; Dobbyn, A. J.; Eckert, F.; Goll, E.; Hampel, C.; Hesselman, A.; Hetzer, G.; Hrenar, T.; Jansen, G.; Köppl, C.; Liu, Y.; Lloyd, A. W.; Mata, R. A.; May, A. J.; McNicholas, S. J.; Meyer, W.; Mura, M. E.; Nicklass, A.; O'Neill, D. P.; Palmieri, P.; Peng, D.; Pflüger, K.; Pitzer, R. M.; Reiher, M.; Shiozaki, T.; Stoll, H.; Stone, A. J.; Tarroni, R.; Thorsteinsson, T.; Wang, M. MOLPRO, Version 2015.1, a Package of Ab Initio Programs. University of Cardiff Chemistry Consultants (UC3):

Cardiff, Wales, UK 2015.
(44) Frisch, M. J.; Trucks, G. W.; Schlegel, H. B.; Scuseria, G. E.; Robb, M. A.; Cheeseman, J. R.; Scalmani, G.; Barone, V.; Mennucci, B.; Petersson, G. A.; Nakatsuji, H.; Caricato, M.; Li, X.; Hratchian, H. P.; Izmaylov, A. F.; Bloino, J.; Zheng, G.; Sonnenberg, J. L.; Hada, M.; Ehara, M.; Toyota, K.; Fukuda, R.; Hasegawa, J.; Ishida, M.; Nakajima, T.; Honda, Y.; Kitao, O.; Nakai, H.; Vreven, T.; Montgomery, J. A., J.; Peralta, J. E.; Ogliaro, F.; Bearpark, M.; Heyd, J. J.; Brothers, E.; Kudin, K. N.; Staroverov, V. N.; Kobayashi, R.; Normand, J.; Raghavachari, K.; Rendell, A. P.; Burant, J. C.; Iyengar, S. S.; Tomasi, J.; Cossi, M.; Rega, N.; Millam, M. J.; Klene, M.; Knox, J. E.; Cross, J. B.; Bakken, V.; Adamo, C.; Jaramillo, J.; Gomperts, R.; Stratmann, R. E.; Yazyev, O.; Austin, A. J.; Cammi, R.; Pomelli, C.; Ochterski, J. W.; Martin, R. L.; Morokuma, K.; Zakrzewski, V. G.; Voth, G. A.; Salvador, P.; Dannenberg, J. J.; Dapprich, S.; Daniels, A. D.; Farkas, Ö.; Foresman, J. B.; Ortiz, J. V.; Cioslowski, J.; Fox, D. J. Gaussian 09 Rev. D01. Gaussian, Inc.: Wallingford, CT 2012.
(45) Raghavachari, K.; Trucks, G. W.; Pople, J. A.; Head-Gordon, M. A Fifth-Order Perturbation Comparison of Electron Correlation Theories. *Chem. Phys. Lett.* **1989**, *157* (6), 479–483 DOI: 10.1016/S0009-2614(89)87395-6.
(46) Watts, J. D.; Gauss, J.; Bartlett, R. J. Coupled-Cluster Methods with Noniterative Triple Excitations for Restricted Open-Shell Hartree–Fock and Other General Single Determinant Reference Functions. Energies and Analytical Gradients. *J. Chem. Phys.* **1993**, *98* (11), 8718–8733 DOI: 10.1063/1.464480.
(47) Shavitt, I.; Bartlett, R. J. *Many – Body Methods in Chemistry and Physics*; Cambridge University Press: Cambridge, 2009.
(48) Peng, D.; Reiher, M. Exact Decoupling of the Relativistic Fock Operator. *Theor. Chem. Acc.* **2012**, *131* (1), 1–20 DOI: 10.1007/s00214-011-1081-y.
(49) Dunning, T. H. Gaussian Basis Sets for Use in Correlated Molecular Calculations. I. The Atoms Boron through Neon and Hydrogen. *J. Chem. Phys.* **1989**, *90* (2), 1007–1023 DOI: 10.1063/1.456153.
(50) Kendall, R. A.; Dunning, T. H.; Harrison, R. J. Electron Affinities of the First-Row Atoms Revisited. Systematic Basis Sets and Wave Functions. *J. Chem. Phys.* **1992**, *96* (9), 6796–6806 DOI: 10.1063/1.462569.
(51) Woon, D. E.; Dunning, T. H. Gaussian Basis Sets for Use in Correlated Molecular Calculations. III. The Atoms Aluminum through Argon. *J. Chem. Phys.* **1993**, *98* (2), 1358–1371 DOI: 10.1063/1.464303.
(52) Dunning, T. H.; Peterson, K. A.; Wilson, A. K. Gaussian Basis Sets for Use in Correlated Molecular Calculations. X. The Atoms Aluminum through Argon Revisited. *J. Chem. Phys.* **2001**, *114* (21), 9244–9253 DOI: 10.1063/1.1367373.
(53) de Jong, W. A.; Harrison, R. J.; Dixon, D. A. Parallel Douglas–Kroll Energy and Gradients in NWChem: Estimating Scalar Relativistic Effects Using Douglas–Kroll Contracted Basis Sets. *J. Chem. Phys.* **2001**, *114* (1), 48 DOI: 10.1063/1.1329891.
(54) Cowan, R. D.; Griffin, D. C. Approximate Relativistic Corrections to Atomic Radial Wave Functions. *J. Opt. Soc. Am.* **1976**, *66* (10), 1010 DOI: 10.1364/JOSA.66.001010.
(55) Martin, R. L. All-Electron Relativistic Calculations on Silver Hydride. An Investigation of the Cowan-Griffin Operator in a Molecular Species. *J. Phys. Chem.* **1983**, *87* (5), 750–754 DOI: 10.1021/j100228a012.
(56) Szalay, P. G.; Bartlett, R. J. Approximately Extensive Modifications of the Multireference Configuration Interaction Method: A Theoretical and Practical Analysis. *J. Chem. Phys.* **1995**, *103* (9), 3600 DOI: 10.1063/1.470243.
(57) Perdew, J.; Burke, K.; Ernzerhof, M. Generalized Gradient Approximation Made Simple. *Phys. Rev. Lett.* **1996**, *77* (18), 3865–3868.
(58) Adamo, C.; Barone, V. Toward Reliable Density Functional Methods without Adjustable Parameters: The PBE0 Model. *J. Chem. Phys.* **1999**, *110* (13), 6158–6170 DOI: 10.1063/1.478522.
(59) Weinhold, F.; Landis, C. R. *Valency and Bonding*; Cambridge University Press: Cambridge, 2005.
(60) Glendening, E. D.; Landis, C. R.; Weinhold, F. NBO 6 . 0 : Natural Bond Orbital Analysis Program. **2013**, 1–9 DOI: 10.1002/jcc.23266.
(61) Montgomery, J. A.; Frisch, M. J.; Ochterski, J. W.; Petersson, G. A. A Complete Basis Set Model Chemistry. VII. Use of the Minimum Population Localization Method. *J. Chem. Phys.* **2000**, *112* (15), 6532–6542 DOI: 10.1063/1.481224.
(62) Cioslowski, J.; Surján, P. R. An Observable-Based Interpretation of Electronic Wavefunctions: Application to "Hypervalent" Molecules. *J. Mol. Struct. THEOCHEM* **1992**, *255*, 9–33 DOI: 10.1016/0166-1280(92)85003-4.

(63) Martin, J. M. L. Heats of Formation of Perchloric Acid, HClO4, and Perchloric Anhydride, Cl2O7. Probing the Limits of W1 and W2 Theory. *J. Mol. Struct. THEOCHEM* **2006**, *771* (1–3), 19–26 DOI: 10.1016/j.theochem.2006.03.035.

(64) Geerlings, P.; De Proft, F.; Langenaeker, W. Conceptual Density Functional Theory. *Chem. Rev.* **2003**, *103* (5), 1793–1874 DOI: 10.1021/cr990029p.